\documentclass[%
reprint,
superscriptaddress,
 amsmath,amssymb,
 aps,
 usenames,dvipsnames,
]{revtex4-2}

\usepackage{dcolumn}
\usepackage{multirow}

\usepackage[utf8]{inputenc}
\usepackage{xcolor}
\usepackage[normalem]{ulem}
\usepackage[final]{graphicx}
\usepackage{bm}
\usepackage{float}

\usepackage[draft=false]{hyperref}
\hypersetup{
    colorlinks,
    citecolor=blue,
    filecolor=blue,
    linkcolor=blue,
    urlcolor=blue
}

\usepackage{natbib}
\newcommand{\ket}[1]{|#1\rangle}
\newcommand{\bra}[1]{\langle#1|}
\newcommand{\braket}[2]{\langle#1|#2\rangle}
\newcommand{\dd}[1]{\mathrm{d} #1}
\newcommand{\abs}[1]{|#1|}
\newcommand{\pqty}[1]{\left(#1\right)}

\begin{document}

\title{Sculpting bosonic states with arithmetic subtractions}

\author{Lin Htoo Zaw}
\thanks{These authors contributed equally to this work.}
\affiliation{Centre for Quantum Technologies, National University of Singapore, 3 Science Drive 2, Singapore 117543}
\author{Zakarya Lasmar}
\thanks{These authors contributed equally to this work.}
\affiliation{Centre for Quantum Technologies, National University of Singapore, 3 Science Drive 2, Singapore 117543}
\author{Chi-Huan Nguyen}
\affiliation{Centre for Quantum Technologies, National University of Singapore, 3 Science Drive 2, Singapore 117543}
\author{Ko-Wei Tseng}
\affiliation{Centre for Quantum Technologies, National University of Singapore, 3 Science Drive 2, Singapore 117543}
\author{Dzmitry Matsukevich}
\affiliation{Centre for Quantum Technologies, National University of Singapore, 3 Science Drive 2, Singapore 117543}
\affiliation{Department of Physics, National University of Singapore, 2 Science Drive 3, Singapore 117542}
\author{Dagomir Kaszlikowski} 
\affiliation{Centre for Quantum Technologies, National University of Singapore, 3 Science Drive 2, Singapore 117543}
\affiliation{Department of Physics, National University of Singapore, 2 Science Drive 3, Singapore 117542}
\author{Valerio Scarani}
\affiliation{Centre for Quantum Technologies, National University of Singapore, 3 Science Drive 2, Singapore 117543}
\affiliation{Department of Physics, National University of Singapore, 2 Science Drive 3, Singapore 117542}

\begin{abstract}
Continuous-variable (CV) encoding allows information to be processed compactly and efficiently on quantum processors. Recently developed techniques such as controlled beam-splitter operations and the near deterministic phonon subtractions make trapped ion systems attractive for exploring CV quantum computing. Here we propose a probabilistic scheme based on the boson sculpting technique for generating multipartite highly entangled states of motional modes of trapped ion systems. We also investigate the effects of decoherence on the fidelity of the generated state by performing numerical simulations with realistic noise parameters. Our work is a step towards generating multipartite continuous-variable entanglement.
\end{abstract}

\maketitle
\date{\today}

\section{Introduction}

Quantum entanglement is a property of a compound system that possesses non-classical correlation between its subsystems. The ability to prepare highly entangled states is crucial for quantum computation. For physical platforms that employ discrete variables, the entanglement between qubits are typically created using short-range interactions induced by bosonic modes. In particular, with the collective motional modes of trapped ions as the quantum bus, the internal degrees of freedom of ions have been entangled with a fidelity significantly above the threshold required for fault-tolerant quantum computation~\cite{Harty, Gaebler, Lucas}. However, with a larger number of qubits, implementing the full control necessary for entangling operations remains challenging due to several technical problems, such as crosstalk, heating, and the overhead of addressing individual qubits.

Alternatively, classical information can be encoded into the eigenstates of continuous-valued operators, such as the motion of trapped ions, the quadratures of an electromagnetic mode, and the spin variables of an atomic ensemble \cite{Es2008,Vlastakis2013,nguyen2021}. In such bosonic systems, a large dimension of the Hilbert space is typically available for the encoding. As less physical resources are required, this makes the quantum computation more efficient.

The bosonic system we study in this paper is the trapped-ion system. The motion of a linearly trapped ion chain is a well-controlled bosonic mode, which allows for deterministic preparation of single-phonon states \cite{Bruzewicz}. There have also been several experiments that report nonlinear gates operating on trapped-ion phonons with good fidelity \cite{Um,squeezed-fock-basis-experiment}, and high-fidelity state reconstruction of the motional state is also possible \cite{state-reconstruction}.

Despite their advantages, it can be more complicated to generate entanglement with bosonic modes, though there have been some early experimental attempts to generate bipartite entanglement~\cite{Brown2011,Wang2016,GaoESwap2019}. Here, we present a scheme based on arithmetic phonon subtraction to prepare a cat state on the motional modes of 4 ions. The proposed method can be generalized to a Hilbert space a higher dimension and more motional modes.

Cat states are evenly populated superpositions of maximally distinguishable states \cite{Leibfried2005}, also referred to as Greenberger–Horne–Zeilinger (GHZ) states. Because they contain genuine multipartite correlations, they are of particular interest for future quantum technologies \cite{Bruzewicz,Omran2019}. This type of entanglement is considered a universal resource for quantum computing \cite{shor1996,Steane05,Knill2005,Gottesman1999}, quantum communication \cite{Zhao2004,Hillery99}, and also for testing the foundations of quantum physics \cite{GHZ90}. Cat states are challenging to prepare because of their sensitivity to decoherence. They induce the so-called \textit{super-decoherence} which can be used as benchmark for robust quantum control \cite{Knill2000,Monz2011,Pezze2018,Pogorelov2021,Mooney_2021}.

In this paper, we propose an experiment for entangling the collective motional modes of 4 trapped ions. The same protocol can be extended to systems with a higher number of ions. The main idea of our approach is to prepare a single phonon in each motional mode, rotate the basis, and subtract half of the phonons. This results in a final state that carries genuine multipartite correlations between the motional modes. Because particle subtraction is the key step for revealing these mode correlations, we refer to this approach as a \textit{sculpting scheme}, which was originally proposed for multimode photonic platforms \cite{sculpting}. Since it is not possible to remove particles from a vacuum, sculpting schemes are probabilistic, with the success probability dependent on the vacuum component of the state prior to the subtraction process.

As quantum entanglement is regarded as a vital resource for quantum technologies, finding new ways for its extraction is of natural interest. Moreover, for many-body systems composed of indistinguishable particles, the intrinsic correlations due the symmetrization constraints are the subject of a rapidly growing interest for their potential applications as a quantum resource \cite{Benatti,Morris,Killoran,Franco}. In the sculpting scheme, these intrinsic correlations are consumed in order to create entanglement between the bosonic modes. Here, we show that the sculpting scheme can also be implemented with trapped ions.

This article is organized as follows: in Section II, we start by presenting the basic idea of using subtraction of indistinguishable bosons for the creation of entanglement between bosonic modes. Then, in Section III, we discuss the necessary operations for adapting this scheme to trapped-ion platforms. In Section IV, we show how these operations can be used to generate entanglement from different initial states. In Section V, we discuss the numerical simulations of an experimental implementation assuming realistic conditions. In the last section, we give an outlook of this work and potential applications. 

\section{Sculpting bosonic GHZ states}

\subsection{Creation of mode entanglement by subtraction}

Let us assume that we have $2n$ bosonic modes. We start by creating a single boson at each mode, corresponding to the symmetric state
\begin{equation}\label{psi0col}
\ket{\mathrm{sym}_{2n}} \equiv \hat{a}^{\dagger}_{1}\hat{a}^{\dagger}_{2}\cdots\hat{a}^{\dagger}_{2n} \ket{\varnothing} = \ket{1_1,1_2,\cdots,1_{2n}} ,
\end{equation}
such that $\hat{a}^{\dagger}_q$ is the creation operator acting on the q$^{th}$ mode and $\ket{\varnothing} = \ket{0_1,\cdots,0_{2n}}$ corresponds to the vacuum for all modes. Here we adopt the shorthand notation $\ket{n_1,0_2,\cdots,0_{2n}}$ which denotes having $n$ particles at the first mode while all remaining modes are empty. 

The initial symmetric state $|\mathrm{sym}_{2n}\rangle$ can be transformed into an entangled state by implementing $n$ successive subtraction operations defined as
\begin{equation}\label{eq:J}
    \hat{\mathcal{J}} = \prod_{j=0}^{n-1} \hat{A}_j,
\end{equation}
where each subtraction operation has the form
\begin{equation}\label{eq:Aj}
    \hat{A}_j = \sum_{p=1}^{n} \hat{a}_p + \sum_{q=n+1}^{2n} e^{i 2(j+q)\pi/n} \hat{a}_q.
\end{equation}

After normalization, we obtain an entangled state of the form \cite{sculpting} 
\begin{equation}
\begin{split}
\ket{\mathrm{GHZ}_{2n}} =& \frac{1}{\sqrt{2}} (\ket{1_1,\cdots,1_n,0_{n+1},\cdots,0_{2n}} \\
&+ (-1)^{n+1} \ket{0_1,\cdots,0_n,1_{n+1},\cdots,1_{2n}}).
\end{split}
\end{equation}

\paragraph*{}
Since our system consists of identical particles distributed over multiple modes, two types of quantum correlations are present: particle entanglement and mode entanglement \cite{Benatti,Morris,Killoran,Franco}. The former is due to the exchange symmetry of the indistinguishable bosons, while the latter corresponds to correlations between different modes. 
In first quantization, the initial state reads
\begin{equation}
\ket{\mathrm{sym}_{2n}} = \frac{1}{\sqrt{2n!}} \sum_{\sigma} \ket{\sigma_{1,2,\cdots,2n}}
\end{equation}
such that $\sigma$ is the set of all possible permutations of having one particle at each mode. Clearly, due to the exchange symmetry, the state above is highly correlated. Since the entangled parties are indistinguishable, the correlations in $\ket{\mathrm{sym}_{2n}}$ are inaccessible. Through the transformation $\ket{\mathrm{sym}_{2n}} \to \ket{\mathrm{GHZ}_{2n}}$, one can notice the following: (i) The amount of particle entanglement is reduced. This is because the number of particles decreases from $2n$ to $n$. Consequently, in first quantization, the set of possible permutations contracts as well. (ii) Mode entanglement of the \textit{GHZ}-type is created.

The choice of $\ket{\text{sym}_{2n}}$ is important, as it has been shown that the initial state must have nonzero \emph{particle} entanglement in order to extract accessible \emph{mode} entanglement using only subtraction operations \cite{Morris}. Some subtlety is involved in quantifying exactly what is exchanged between the two types of entanglement, which is briefly discussed in appendix~\ref{apd:ME-PE-Complications}. In short, by performing $n$ subtractions, some of the inaccessible particle entanglement is consumed in order to create accessible entanglement between bosonic modes \cite{sculpting}.

\subsection{Arithmetic subtraction}

We have just introduced boson sculpting with the usual ladder operators $\hat{a}^{\dagger} = \sum_{n=0}^{\infty} \sqrt{n+1}|n+1\rangle\langle n|$ and $\hat{a}^{} = \sum_{n=0}^{\infty} \sqrt{n+1}|n\rangle\langle n+1|$. They can be implemented in trapped-ion systems using beam-splitting operations \cite{Toyoda} but require additional ancillary motional modes \cite{sculpting}, i.e.~having additional trapped ions will be necessary.

In this paper, we shall rather focus on the so-called \textit{arithmetic operations}  \cite{Um}
\begin{eqnarray}\label{}
    \hat{S}^{\dagger} &= \sum_{n=0}^{\infty} |n+1\rangle\langle n|, \\
    \hat{S}^{} &= \sum_{n=0}^{\infty} |n\rangle\langle n+1| \;.\label{arthmsub}
\end{eqnarray}
The arithmetic operations are referred to as ``\textit{near} deterministic" \cite{Um} because the only deviation from unitarity is due to $\hat{S}\ket{0}=0$: that is, while $\hat{S}\hat{S}^\dagger=\mathbb{I}$ holds, one finds $\hat{S}^\dagger\hat{S}=\mathbb{I}-\ket{0}\bra{0}$. In particular, the subtraction $S$ preserves the scalar product of all pairs of states (hence, the norm of all states) that do not have a vacuum component.

In ion traps, arithmetic operations can be implemented without ancillary ions via \textit{adiabatic schemes} \cite{Bergmann,Vitanov,Gebert}. If the adiabatic passage is performed slowly enough, and for a long enough time period, one can be certain that the populations have been completely transferred from each state $|n\rangle$ to $|n+1\rangle$ while maintaining the coherences \cite{Um}.

In principle, the inclusion of arithmetic subtraction to the set of Gaussian operations performed in a trapped-ion system allows for universal state preparation \cite{universal-CV}. However, there is no general method available to find the sequence of arithmetic subtraction and Gaussian operations required to prepare an arbitrary state. As such, we focus on reporting the exact state preparation of the $\ket{\text{GHZ}_{4}}$ state, which has a known use in quantum computation \cite{byzantine}.

\subsection{Bosonic sculpting with trapped ions}
\label{ss:scenarios}

\begin{figure*}[htp]
(a)
\begin{center}
   \includegraphics[width=0.73\textwidth]{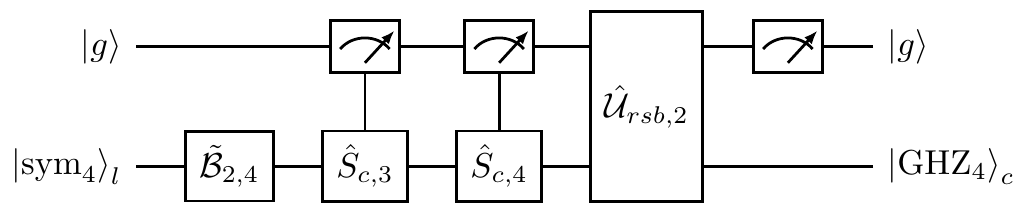}
\end{center}
(b)
\begin{center}
   \includegraphics[width=\textwidth]{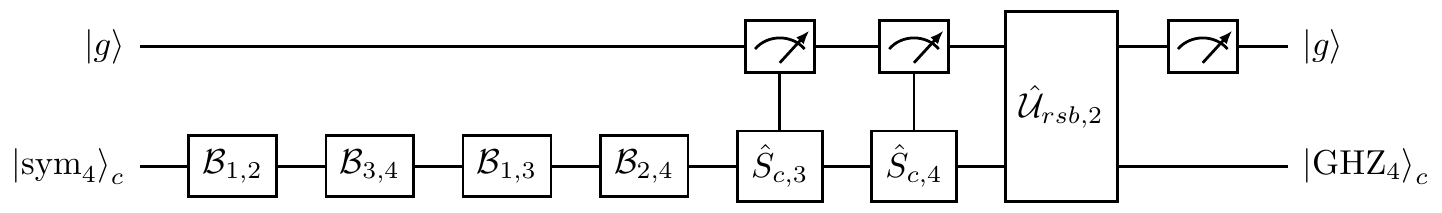}
\end{center}
    \caption{Circuit for creation of entanglement between the collective modes of 4 trapped ions. (a)  The initial state is prepared in the local basis. Following the convention (\ref{preconvension}), the beam-splitting gate $\tilde{\mathcal{B}}_{4,2}$ has the parameters $\phi = -\pi/2$ and $\theta = 2 \lambda - \pi/2$. (b) The initial state is prepared in the collective basis. All the gates $\mathcal{B}_{p,q}$ are 50-50 beam-splitting operations, i.e. $\phi = -\pi/2$ and $\theta = \pi/2$. The gate $\hat{S}_{c,k}$ correspond to an arithmetic subtraction from the $k^{th}$ collective mode. The implementation of the arithmetic subtraction process involves post-selecting the ground state of the spin degree of freedom, as detailed in subsection~\ref{ss:arithmetic-subtraction-implementation}. The remaining gates correspond to a red sideband transition, followed by another post-selection of the ground state, to obtain exactly the target state $\ket{\phi_4}_c$. }\label{circuitab}
        \label{fig:cuircuitsab}
\end{figure*}
In trapped-ion systems, we can address the motional modes of the ions in two bases: the local basis, and the collective basis. The local basis, whose states and operators we will denote with a subscript $l$, corresponds to the motion of the $l$\textsuperscript{th} ion. The collective basis, whose states and operators we will denote with a subscript $c$, refers to the normal modes of the collective motion of the ion chain.

In this proposal, we will show how to create the target state $|\mathrm{GHZ}_4\rangle_{c}$, entangled in the collective basis, starting from an initial state prepared either in the local basis $\ket{\mathrm{sym}_4}_l$ or the collective one $\ket{\mathrm{sym}_4}_c$. The gate sequence for both scenarios is shown in Fig.~\ref{circuitab}. The first scenario (subsection \ref{ss:scen1}) requires fewer gates, but is harder to implement faithfully because of the phonon-hopping between the local modes. The second scenario (subsection \ref{ss:scen2}) avoids this problem by working solely in the collective basis, but requires an implementation time that is almost thrice as long, and is thus more affected by noise.

\subsubsection{Scenario with individual addressing}
\label{ss:scen1}

For the appropriate choice of the Paul-trap parameters, the Hamiltonian describing the trapped ions will have eigenvectors that coincide with the so-called collective modes of motion \cite{Leibfried}. In the case of 4 trapped ions, these modes are related to the local ones via the relations \cite{James}:
\begin{eqnarray}\label{cbasis}
    \begin{array}{lll}
        &\hat{a}_{c,1} =& \frac{1}{2}(\hat{a}^{}_{l,1} + \hat{a}^{}_{l,2} + \hat{a}^{}_{l,3} + \hat{a}^{}_{l,4}) \;, \\[1.5ex]
        &\hat{a}_{c,2} =& \frac{1}{\sqrt{2}}(-\mathcal{C} \hat{a}^{}_{l,1} -  \mathcal{S} \hat{a}^{}_{l,2} + \mathcal{S} \hat{a}^{}_{l,3} +  \mathcal{C} \hat{a}^{}_{l,4}) \;,\\[1.5ex]
        &\hat{a}_{c,3} =& \frac{1}{2}(\hat{a}^{}_{l,1} - \hat{a}^{}_{l,2} - \hat{a}^{}_{l,3} + \hat{a}^{}_{l,4}) \;, \\[1.5ex]
        &\hat{a}_{c,4}=&  \frac{1}{\sqrt{2}}( - \mathcal{S} \hat{a}^{}_{l,1} + \mathcal{C} \hat{a}^{}_{l,2}  -\mathcal{C} \hat{a}^{}_{l,3} +\mathcal{S} \hat{a}^{}_{l,4}) \;,
    \end{array}
\end{eqnarray}
such that the operator $ \hat{a}^{\dagger}_{c,k} $ $(\hat{a}^{}_{c,k})$ creates (annihilates) one phonon in the $k^{th}$ collective mode, while the subscript $l$ labels the local modes. By including the leading order of the Coulomb interaction terms, the numerical diagonalization results in $\mathcal{C} = \cos[\lambda]$ and $\mathcal{S} = \sin[\lambda]$ with $\lambda = 0.306277$  (see Table 2 of the reference \cite{James}). 

The initial state is prepared in the local basis $\ket{\mathrm{sym}_4}_l$, and the first gate to act on it is a beam-splitting operation between the $2^{nd}$ and $4^{th}$ collective modes $\tilde{\mathcal{B}}_{2,4} \equiv \hat{B}_{2,4}(2\lambda - \frac{\pi}{2}, -\frac{\pi}{2})$. For the sake of convenience, we define here the following ion-trap beam-splitting convention:
\begin{equation}\label{preconvension}
    \hat{B}_{j,k}(\theta,\phi):
    \Bigg\{
    \begin{array}{rll}
      \hat{a}_{j} & \rightarrow & \cos[\theta /2]\hat{a}_{j} - i e^{i \phi} \sin[\theta /2]\hat{a}_{k} \;,\\
      \hat{a}_{k} & \rightarrow & \cos[\theta /2]\hat{a}_{k} - i e^{-i \phi} \sin[\theta /2]\hat{a}_{j} \;.
    \end{array}
\end{equation}
The implementation details of these beam-splitting gates will be discussed later (see subsection \ref{sec:BS}). Using the above definitions in (\ref{preconvension}) and (\ref{cbasis}), we can write 
\begin{equation}
    \begin{split}
        \tilde{\mathcal{B}}_{2,4}\ket{\mathrm{sym}_4}_l =\; & \frac{1}{16} \big(\hat{a}^{\dagger}_{c,1} + \hat{a}^{\dagger}_{c,2} + \hat{a}^{\dagger}_{c,3} + \hat{a}^{\dagger}_{c,4} \big) \\ 
        &\times \big(\hat{a}^{\dagger}_{c,1} + \hat{a}^{\dagger}_{c,2} - \hat{a}^{\dagger}_{c,3} - \hat{a}^{\dagger}_{c,4} \big)  \\
        &\times \big(\hat{a}^{\dagger}_{c,1} - \hat{a}^{\dagger}_{c,2} + \hat{a}^{\dagger}_{c,3} - \hat{a}^{\dagger}_{c,4} \big) \\ 
        &\times\big(\hat{a}^{\dagger}_{c,1} - \hat{a}^{\dagger}_{c,2} - \hat{a}^{\dagger}_{c,3} + \hat{a}^{\dagger}_{c,4} \big)\ket{\varnothing} \;. \label{befores3s4}
    \end{split}
\end{equation}
By expanding the product above, one can easily see that after subtracting one particle each from the $3^{rd}$ and $4^{th}$ collective modes, only two terms will survive
\begin{equation}\label{afters3s4}
    \hat{S}_{c,3}\hat{S}_{c,4} \tilde{\mathcal{B}}_{2,4} \ket{\mathrm{sym}_{4}}_{l} \xrightarrow{\mathrm{normalization}} \ket{\psi_{\mathrm{f}}} \;,
\end{equation}
such that 
\begin{equation}
    \ket{\psi_{\mathrm{f}}} = \frac{1}{\sqrt{5}} \left(  2\ket{1_1,1_2,0_3,0_4}_c  - \ket{0_1,0_2,1_3,1_4}_c  \right)  \;.
\end{equation}
Comparing this to the target state, we obtain the ideal fidelity $\abs{\braket{\psi_{\mathrm{f}}}{\text{GHZ}_4}_c}^2 = 3/\sqrt{10} \approx 0.949$. With arithmetic subtractions, we do not obtain $\ket{\text{GHZ}_4}_c$ with unit fidelity: $\hat{S}_{c,k}$ does not give rise to the usual factors of $\sqrt{n}$ like the ordinary annihilation operator $\hat{a}^{}_{c,k}$, so the two constituent states making up the superposition in $\ket{\psi_{\mathrm{f}}}$ are not evenly populated. While this state is no longer separable, it is not yet maximally entangled. In order to increase the correlations between the motional modes of the ions, we need to evolve the state as 
 \begin{equation}\label{eq:rsb-trick}
    \begin{split}
    |\psi_{\mathrm{f}} \rangle \rightarrow \frac{1}{\sqrt{5}} \Bigg[ & 2 \Big(\frac{1}{2}\ket{g}\ket{1_1,1_2,0_3,0_4}_c \\
    &+ \frac{\sqrt{3}}{2}\ket{e}\ket{1_1,0_2,0_3,0_4}_c \Big)  \\  
    &- \ket{g}\ket{0_1,0_2,1_3,1_4}_c \Bigg] \;.
    \end{split}
\end{equation}
Here, $\ket{g}$/$\ket{e}$ refers to the internal spin state of the ion addressed by the Raman lasers---see subsection~\ref{ss:Hamiltonian} for more details. The transformation above can be implemented using the red sideband transition on the 2$^{nd}$ collective mode, which will be discussed in subsection \ref{sec:RSB}. As a final step, we post-select the ground state $|g\rangle$ to get, after normalization,
\begin{equation}
    |\mathrm{GHZ}_{4}\rangle_{c}  =  \frac{1}{\sqrt{2}}\ket{g} \left( \ket{1_1,1_2,0_3,0_4}_c - \ket{0_1,0_2,1_3,1_4}_c \right) \;.
\end{equation}
All beam-splitting operations considered in this article are between collective modes. However, if the initial state is prepared in the local basis as $\ket{\mathrm{sym}_4}_l = \hat{a}^{}_{l,1}\hat{a}^{}_{l,2}\hat{a}^{}_{l,3}\hat{a}^{}_{l,4}\ket{\varnothing}$, one will need to use the relations (\ref{cbasis}) to appropriately compute the output state. If the initial state is prepared in the collective basis $\ket{\mathrm{sym}_4}_c = \hat{a}^{}_{c,1}\hat{a}^{}_{c,2}\hat{a}^{}_{c,3}\hat{a}^{}_{c,4}\ket{\varnothing}$, then the transformation in (\ref{preconvension}) can be used in a straightforward manner.

\subsubsection{Scenario without individual addressing}
\label{ss:scen2}

In the second scenario, the ions are not addressed individually. The initial state is prepared in the collective basis $\ket{\mathrm{sym}_4}_{c}$. First, let us define a sequence of 50-50 beam-splitting operations as follows
\begin{equation}\label{eq:bs-seq-without-ia}
    C^{\dagger} \equiv \hat{B}_{2,4}\left(\scriptstyle\frac{\pi} {2},-\frac{\pi}{2}\right)\hat{B}_{1,3}\left(\scriptstyle\frac{\pi} {2},-\frac{\pi}{2}\right)\hat{B}_{3,4}\left(\scriptstyle\frac{\pi} {2},-\frac{\pi}{2}\right)\hat{B}_{1,2}\left(\scriptstyle\frac{\pi} {2},-\frac{\pi}{2}\right) \;.
\end{equation}
Using the definition \eqref{preconvension}, one can easily verify that the operation $\hat{B}_{j,k}\left(\scriptstyle\frac{\pi} {2},-\frac{\pi}{2}\right)$ is a 50-50 beam-splitting gate between the j$^{th}$ and k$^{th}$ modes. If we apply sequence above to the initial state, we get
\begin{equation}\label{eq:befores3s4-without-ia}
\begin{split}
     C^{\dagger}\ket{\mathrm{sym}_4}_c = \; &\frac{1}{16} \big(\hat{a}^{\dagger}_{c,1} + \hat{a}^{\dagger}_{c,2} + \hat{a}^{\dagger}_{c,3} + \hat{a}^{\dagger}_{c,4} \big) \\
&\times\big(\hat{a}^{\dagger}_{c,1} + \hat{a}^{\dagger}_{c,2} - \hat{a}^{\dagger}_{c,3} - \hat{a}^{\dagger}_{c,4} \big) \\
&\times \big(\hat{a}^{\dagger}_{c,1} - \hat{a}^{\dagger}_{c,2} + \hat{a}^{\dagger}_{c,3} - \hat{a}^{\dagger}_{c,4} \big) \\
&\times \big(\hat{a}^{\dagger}_{c,1} - \hat{a}^{\dagger}_{c,2} - \hat{a}^{\dagger}_{c,3} + \hat{a}^{\dagger}_{c,4} \big)\ket{\varnothing} \;.
\end{split}
\end{equation}
This is exactly the state (\ref{befores3s4}) that we previously obtained after the beam-splitting gate $\tilde{\mathcal{B}}_{2,4}$. Therefore, to reach the target state $\ket{\mathrm{GHZ}_4}_c$, one has to perform the same sequence of operations as before (subtract from the $3^{rd}$ and $4^{th}$ modes, drive the red sideband transition of the $2^{nd}$ mode, and finally post-select the ground state of the internal degree of freedom).

In general, this scheme can be extended to a $\ket{\text{GHZ}_{2n}}$ state for any $n$, with and without the red sideband correction. We discuss the general case in appendix~\ref{apd:general-GHZ-n}. For this experimental proposal, we focus only on the simplest nontrivial case of $n=2$.

\subsubsection{Success probabilities}
Due to the post-selective measurements, required for both the arithmetic subtraction and the corrective red sideband operation, this scheme is probabilistic. The success probabilities are $5/16 = 31.25\%$ without the red sideband correction, and $1/8 = 12.5\%$ with the red sideband correction. These probabilities are the same for the scenario with and without individual addressing.

In trapped-ion systems, the initial state $\ket{\text{sym}_{2n}}$ is deterministically prepared from the motional ground state with the red sideband transition, as will be covered in section~\ref{sec:RSB}. The measurement of the internal state of the ion is also highly efficient, with a state detection fidelity of $\gtrsim 0.999$ \cite{hifi-state-detection}. Therefore, the success probability of preparing $\ket{\text{GHZ}_{2n}}$ is determined solely by the success probability of the subtraction sequence and the red sideband correction, as reported above.

\section{Bosonic modes of trapped ions}

\subsection{Hamiltonian of the system}\label{ss:Hamiltonian}

For the concrete proposal, we refer to the experimental setup reported in \cite{Gan,Nguyen,nguyen2021}: a system of four ions in a linear Paul trap, with trap frequencies $\omega_z \ll \omega_x,\omega_y$. We address the motional degree of freedom of the ions along the $x$ direction and the internal spin state of the last (fourth) ion.
The last ion is chosen since it can participate in all motional modes. The Hamiltonian of the system takes the form \cite{James,Marquet}
\begin{equation}
\hat{H}_0 = \frac{\hbar\omega_0}{2}\hat{\sigma}_z + \sum_{j=1}^4 \hbar\nu_j\left(\hat{a}^{\dagger}_{c,j}\hat{a}^{}_{c,j}+\case{1}{2}\right).
\end{equation}
Here, $\nu_j$ and $\hat{a}_{c,j}$ are the frequency and annihilation operator of the collective mode $j$, $\omega_0$ is the carrier transition frequency, and $\hat{\sigma}_z = \vert e \rangle\!\langle e \vert - \vert g \rangle\!\langle g \vert$, where $\vert e \rangle$ and $\vert g \rangle$ are the excited and ground states of the spin.

The system is driven by a pair of lasers that couple the internal states of the fourth ion to the collective motional modes via Raman transitions. We denote the laser frequencies as $\omega_{L,1}$ and $\omega_{L,2}$,  and their phases as $\phi_{1}$ and $\phi_{2}$.
In the rotating frame of the free Hamiltonian $\hat{H_0}$, and after taking the rotating-wave approximation, the ion-laser coupling is governed by the interaction Hamiltonian \cite{Shen}
\begin{equation}\label{eq:full-hamiltonian}
        \hat{\mathcal{H}}_{\mathrm{I}}(t) = \sum_{k = 1}^5 \hat{\mathcal H}_{\mathrm{I},k} (t),
\end{equation}
where
\begin{widetext}
\begin{equation}\label{eq:HI1}
    \hat{\mathcal H}_{\mathrm{I},1} (t) = 
            \sum_{l=1}^{2}\frac{\hbar g_l}{2}\left[
                \bigg(
                    \hat{\sigma}_+ - \sum_{j=1}^4 \eta_j^2 (
                        \hat{a}_{c,j}^\dagger\hat{a}_{c,j} + \case{1}{2}
                    )
                \bigg)
                    e^{i\phi_l}
                    e^{-i\delta_l t} +
                \mathrm{H.c.}
            \right]\;,
\end{equation}
\begin{equation}\label{eq:HI2}
    \hat{\mathcal H}_{\mathrm{I},2} (t) = \sum_{j=1}^{4} \left\{
            \sum_{l=1}^2\frac{\hbar g_l\eta_j}{2}\bigg[
                \hat{\sigma}_+\hat{a}^\dagger_{c,j}
                    e^{i(\phi_l+\frac{\pi}{2})}
                    e^{-i(\delta_l-\nu_j)t} +
                \mathrm{H.c.}
            \bigg]
        \right\}\;,
\end{equation}
\begin{equation}\label{eq:HI3}
    \hat{\mathcal H}_{\mathrm{I},3} (t) = \sum_{j=1}^{4} \left\{
            \sum_{l=1}^2\frac{\hbar g_l\eta_j}{2}\bigg[
                \hat{\sigma}_+\hat{a}_{c,j}
                    e^{i(\phi_l+\frac{\pi}{2})}
                    e^{-i(\delta_l+\nu_j)t} +
                    \mathrm{H.c.}
            \bigg]
        \right\}\;,
\end{equation}
\begin{equation}\label{eq:HI4}
    \hat{\mathcal H}_{\mathrm{I},4} (t) = \sum_{j=1}^4\sum_{k=j+1}^4 \Bigg\{
            -\sum_{l=1}^2\frac{\hbar g_l\eta_j\eta_k}{2}\bigg[
                \hat{\sigma}_+ \hat{a}_{c,j}\hat{a}^\dagger_{c,k} e^{i\phi_l}
                    e^{-i(\delta_l-(\nu_k-\nu_j))t} +
                \hat{\sigma}_- \hat{a}_{c,j}\hat{a}^\dagger_{c,k} e^{-i\phi_l}
                    e^{i(\delta_l+(\nu_k-\nu_j))t} + \mathrm{H.c.}
            \bigg]
        \Bigg\}\;,
\end{equation}
\begin{equation}\label{eq:HI5}
\begin{split}
    \hat{\mathcal H}_{\mathrm{I},5} (t) =& \sum_{j=1}^4 \Bigg\{
            -\sum_{l=1}^2\frac{\hbar g_l\eta_j^2}{4}\bigg[
                \hat{\sigma}_+ \hat{a}_{c,j}^{\dagger2}
                    e^{i\phi_l}
                    e^{-i(\delta_l-2\nu_j)t} +
                \hat{\sigma}_- \hat{a}_{c,j}^{\dagger2}
                    e^{-i\phi_l}
                    e^{i(\delta_l+2\nu_j)t} +
                \mathrm{H.c.}
            \bigg]
        \Bigg\}\;.
\end{split}
\end{equation}
\end{widetext}
Here, $\{\hat{a}_{c,j}\}_{j=1}^4$ are annihilation operators acting on the collective motional modes, while $\{\eta_{j}\}_{j=1}^4$ are the Lamb-Dicke parameters.

The terms in each brace in equations (\ref{eq:HI1}-\ref{eq:HI5}) correspond to a resonant frequency. They be addressed by adjusting the effective laser detunings $\delta_1:=\omega_{L,1}-\omega_0$ and $\delta_2:=\omega_{L,2}-\omega_0$ such that the corresponding terms in the Hamiltonian become time independent, while the contributions of the rapidly oscillating off-resonant terms become negligible.

By addressing different resonant frequencies, the various quantum operations required for the boson sculpting scheme can be performed. We list these quantum operations, and the laser parameters required to perform them, in Table \ref{table:detunings-transformations}.

\begingroup
\renewcommand*{\arraystretch}{2}
\begin{table*}
\caption{\label{table:detunings-transformations}Quantum operations with their corresponding laser parameters and state transformations. For the displacement ($\hat{D}_j$) and beam-splitting ($\hat{B}_{j,k}$) operations, the spin state of the system is set to $\ket{-} = \frac{1}{\sqrt{2}}(\ket{e}-\ket{g})$.}
\begin{ruledtabular}
\begin{tabular}{ccccccc}
    & $\delta_1$ & $\delta_2$ & $\phi_1$ & $\phi_2$ & $\theta/\int_0^tdt' g(t')$
        & {Transformations Performed} \\ \hline
\multirow{2}{*}{$\hat{U}_{\mathrm{carr}}$}
    & \multirow{2}{*}{0} & \multirow{2}{*}{0} & \multirow{2}{*}{$-\phi$} & \multirow{2}{*}{$-\phi$} & \multirow{2}{*}{$2$}
                        & $ \ket{g} \to \cos(\case{\theta}{2})\ket{g} - i e^{i\phi} \sin(\case{\theta}{2})\ket{e} $ \\
    &
        &
            &
                &
                    &
                        & $\ket{e} \to
                \cos(\case{\theta}{2})\ket{e} -
                i e^{-i\phi} \sin(\case{\theta}{2})\ket{g}$ \\ \hline
\multirow{2}{*}{$\hat{U}_{\mathrm{rsb},j}$}
    & \multirow{2}{*}{$-\nu_j$} & \multirow{2}{*}{$-\nu_j$} & \multirow{2}{*}{$-\phi-\frac{\pi}{2}$} & \multirow{2}{*}{$-\phi-\frac{\pi}{2}$} & \multirow{2}{*}{$2\eta_j$}
        & $ \ket{e}\ket{n}_{c,j} \to
            \cos(\case{\theta\sqrt{n+1}}{2})
                {\ket{e}\ket{n}_{c,j}} -
            i e^{i\phi} \sin(\case{\theta\sqrt{n+1}}{2})
                {\ket{g}\ket{n+1}_{c,j}} $ \\
    &
        &
            &
                &
                    &
                        &   ${\ket{g}\ket{n+1}_{c,j}} \to
            \cos(\case{\theta\sqrt{n+1}}{2})
                {\ket{g}\ket{n+1}_{c,j}} -
            i e^{-i\phi} \sin(\case{\theta\sqrt{n+1}}{2})
                {\ket{e}\ket{n}_{c,j}}$ \\ \hline  
\multirow{2}{*}{$\hat{U}_{\mathrm{bsb},j}$}
    & \multirow{2}{*}{$\nu_j$} & \multirow{2}{*}{$\nu_j$} & \multirow{2}{*}{$-\phi-\frac{\pi}{2}$} & \multirow{2}{*}{$-\phi-\frac{\pi}{2}$} & \multirow{2}{*}{$2\eta_j$}
        & $ \ket{e}\ket{n+1}_{c,j} \to
            \cos(\case{\theta\sqrt{n+1}}{2})
                {\ket{e}\ket{n+1}_{c,j}} -
            i e^{i\phi} \sin(\case{\theta\sqrt{n+1}}{2})
                {\ket{g}\ket{n}_{c,j}} $ \\
    &
        &
            &
                &
                    &
                        &   ${\ket{g}\ket{n}_{c,j}} \to
            \cos(\case{\theta\sqrt{n+1}}{2})
                {\ket{g}\ket{n}_{c,j}} -
            i e^{-i\phi} \sin(\case{\theta\sqrt{n+1}}{2})
                {\ket{e}\ket{n+1}_{c,j}}$ \\ \hline  
$\hat{D}_{j}$
    & $\nu_j$ & $-\nu_j$ & $-\phi$ & $-\phi-\pi$ & $\eta_j /2$
        & $\hat{a}_{c,j} \to \hat{a}_{c,j} - \theta e^{i\phi}$ \\ \hline  
\multirow{2}{*}{$\hat{B}_{j,k}$}
    & \multirow{2}{*}{$\nu_j-\nu_k$} & \multirow{2}{*}{$\nu_k-\nu_j$} & \multirow{2}{*}{$\pi-\phi$} & \multirow{2}{*}{$\phi-\pi$} & \multirow{2}{*}{$\eta_j\eta_k$}
        & $ \hat{a}_{c,j} \to
                \cos(\case{\theta}{2}) \hat{a}_{c,j} -
                    i e^{i\phi} \sin(\case{\theta}{2}) \hat{a}_{c,k} $ \\
    &
        &
            &
                &
                    &
                        &   $\hat{a}_{c,k} \to
                \cos(\case{\theta}{2}) \hat{a}_{c,k} -
                    i e^{-i\phi} \sin(\case{\theta}{2}) \hat{a}_{c,j}$
\end{tabular}
\end{ruledtabular}
\end{table*}
\endgroup

\subsection{Basic operations}

In the following sections, the coupling strengths of the Raman lasers are taken to be equal and time dependent: that is, $g_1=g_2=g(t)$. Furthermore, all relevant Hamiltonians will be found to be in the form $\hat{H}(t) = \hbar g(t)\hat{G}$ for some time-independent operator $\hat{G}$. The time evolution operator of a system evolving under such a Hamiltonian is given by $\hat{U}(t) = \exp(-i\int_0^{t}dt'g(t')\hat{G})$.

\subsubsection{Carrier Transition}
The carrier transition is zeroth order with respect to the Lamb Dicke parameter, which can be performed by setting the laser detunings to $\delta_1=\delta_2=0$, and phases to $\phi_1=\phi_2=-\phi$, resulting in the Hamiltonian
\begin{equation}
\begin{split}
\hat{H}_{\mathrm{carr}}(t)
=& \hbar \frac{2g(t)}{2}\left( \hat{\sigma}_+ e^{-i\phi} + \hat{\sigma}_- e^{i\phi} \right) \;, \\
=& \hbar \frac{2g(t)}{2}\left( \cos\phi\,\hat{\sigma}_x + \sin\phi\,\hat{\sigma}_y \right) \;.
\end{split}
\end{equation}
The time evolution operator of this Hamiltonian, $\hat{U}_{\mathrm{carr}}(\theta,\phi)$, with $\theta = 2\int_0^t\,dt' g(t')$, is immediately recognizable as a rotation of the spin state
\begin{equation}
\def\arraystretch{1.25}\setlength\arraycolsep{0.5ex}
\begin{array}{rl}
    \hat{U}_{\mathrm{carr}}(\theta,\phi)\vert{g}\rangle &=
        \cos(\case{\theta}{2})\vert{g}\rangle -
        i e^{i\phi} \sin(\case{\theta}{2})\vert{e}\rangle \;,\\
    \hat{U}_{\mathrm{carr}}(\theta,\phi)\vert{e}\rangle &=
        \cos(\case{\theta}{2})\vert{e}\rangle -
        i e^{-i\phi} \sin(\case{\theta}{2})\vert{g}\rangle \;.
\end{array}
\end{equation}
This transition is required for the preparation of motional states in conjunction with the red sideband transitions, and for setting the spin to the correct state to perform operations on the motional degree of freedom.

\subsubsection{Red Sideband Transition\label{sec:RSB}}

The red sideband transition is a first order operation that can be performed by setting the laser detunings to $\delta_1=\delta_2=-\nu_j$, and phases to $\phi_1=\phi_2=-\phi-\frac{\pi}{2}$, such that
\begin{eqnarray}\label{eq:rsb-hamiltonian}
\hat{H}_{\mathrm{rsb},j}(t)
&= \hbar\frac{2g(t)\eta_j}{2}\left(
    \hat{\sigma}_+\hat{a}_{c,j} e^{-i\phi} + \hat{\sigma}_-\hat{a}^\dagger_{c,j} e^{i\phi}
\right) \;.
\end{eqnarray}
The time evolution operator $U_{\mathrm{rsb},j}(\theta,\phi)$, with $\theta = 2\eta_j\int_0^t\,dt' g(t')$, results in Rabi oscillations between the states ${\vert e \rangle\vert n \rangle_{c,j}}\leftrightarrow{\vert g \rangle\vert n+1 \rangle_{c,j}}$, where
\begin{equation}\label{eq:rsb-definition1}
\begin{split}
    U_{\mathrm{rsb},j}(\theta,\phi)
        {\vert e \rangle\vert n \rangle_{c,j}} =&
    \cos(\scriptstyle\case{\theta\sqrt{n+1}}{2})
        {\vert e \rangle\vert n \rangle_{c,j}} \\
        & - i e^{i\phi} \sin(\scriptstyle\frac{\theta\sqrt{n+1}}{2})
        {\vert g \rangle\vert n+1 \rangle_{c,j}} \;,
\end{split}
\end{equation}

\begin{equation}\label{eq:rsb-definition2}
\begin{split}
    U_{\mathrm{rsb},j}(\theta,\phi)
        {\vert g \rangle\vert n+1 \rangle_{c,j}} =&
    \cos(\scriptstyle\case{\theta\sqrt{n+1}}{2})
        {\vert g \rangle\vert n+1 \rangle_{c,j}} \\
        &-i e^{-i\phi} \sin(\scriptstyle\case{\theta\sqrt{n+1}}{2})
        {\vert e \rangle\vert n \rangle_{c,j}} \;.
\end{split}
\end{equation}
At this point, one can easily verify the transformation \eqref{eq:rsb-trick} by substituting the values $\theta=2\pi/3$ and $\phi=\pi/2$ into (\ref{eq:rsb-definition1}-\ref{eq:rsb-definition2}). 

Also, this transition can be used to prepare the initial state required for the boson sculpting scheme. Consider the state $\vert{g}\rangle\vert{0}\rangle_{c,j}$. If a carrier transition is performed, followed by a red sideband transition on mode $j$, with $\theta=\pi,\phi=\pi/2$ for both transitions, the state undergoes the evolution \[\vert{g}\rangle\vert{0}\rangle_{c,j}\xrightarrow{\mathrm{carr}}\vert{e}\rangle\vert{0}\rangle_{c,j}\xrightarrow{\mathrm{rsb,j}}\vert{g}\rangle\vert{1}\rangle_{c,j} \;.\] 
Repeating this sequence of transitions for all modes $j$, we would have
\[ \ket{g}\ket{0_1,0_2,0_3,0_4}_c \to \ket{g}\ket{1_1,1_2,1_3,1_4}_c = \ket{\mathrm{sym}_4}_c \;.\]

\subsubsection{Blue Sideband Transition}
The blue sideband transition is also a first order operation, which can be performed by setting $\delta_1=\delta_2=\nu_j$ and $\phi_1=\phi_2=-\phi-\frac{\pi}{2}$, such that
\begin{equation}
\hat{H}_{\mathrm{bsb},j}(t)
= \hbar\frac{2g(t)\eta_j}{2}\left(
    \hat{\sigma}_+\hat{a}^{\dagger}_{c,j} e^{-i\phi} + \hat{\sigma}_-\hat{a}_{c,j} e^{i\phi}
\right) \;.
\end{equation}
The time evolution $U_{\mathrm{bsb},j}(\theta,\phi)$, with $\theta = 2\eta_j\int_0^t\,dt' g(t')$, carries out the transformation
\begin{equation}
\begin{split}
    U_{\mathrm{bsb},j}(\theta,\phi)
        {\ket{e}\ket{n+1}_{c,j}} =&
    \cos(\scriptstyle\case{\theta\sqrt{n+1}}{2})
        {\vert e \rangle\vert n+1 \rangle_{c,j}} \\
        &\quad - i e^{i\phi} \sin(\scriptstyle\case{\theta\sqrt{n+1}}{2})
        {\vert g \rangle\vert n \rangle_{c,j}} \;, \nonumber 
\end{split}
\end{equation}
\begin{equation}
\begin{split}
    U_{\mathrm{bsb},j}(\theta,\phi)
        {\ket{g}\ket{n}_{c,j}} =&
    \cos(\scriptstyle\case{\theta\sqrt{n+1}}{2})
        {\vert g \rangle\vert n \rangle_{c,j}} \\
        &\quad -i e^{-i\phi} \sin(\scriptstyle\case{\theta\sqrt{n+1}}{2})
        {\vert e \rangle\vert n+1 \rangle_{c,j}} \;. \nonumber
\end{split}
\end{equation}
Generally, operations involving the red sideband transition can also be performed with the blue sideband by first performing a $\pi$-rotation on the spin. In this paper, we address the red sideband instead of the blue sideband wherever required.

\subsection{Composite operations}
In the previous subsection, we presented the 3 basic transitions necessary to manipulate the internal and motional degrees of freedom of the trapped ions. In the following, we will show how these transitions can be used to implement Gaussian gates such as displacement and beam-splitting operations. While the former will be necessary for the tomography of the system, the latter will be useful for preparing the ions in the appropriate basis. Also, we can implement non-Gaussian gates, namely, arithmetic operations, which will be used for subtracting phonons from our system.

\subsubsection{Displacement Operation}
This is a first order operation performed by driving both sideband transitions simultaneously with $\delta_1=-\delta_2=\nu_j$, $\phi_1=-\phi$, and $\phi_2=\phi-\pi$, such that
\begin{equation}
\hat{H}_{D,j}(t) = -i\hbar\sigma_{x} \frac{g(t)\eta_j}{2}\left( \hat{a}^{\dagger}_{c,j} e^{i\phi} -\hat{a}_{c,j} e^{-i\phi} \right) \;.
\end{equation}
By setting the spin state of the system to $\ket{-} := \frac{1}{\sqrt{2}}\left(\vert e \rangle-\vert g \rangle\right)$ with the carrier transition, the time evolution operator corresponds to a displacement operation $\hat{D}_j(\theta,\phi) := \exp\left(\theta(\hat{a}_{c,j}^\dagger e^{i\phi}-\hat{a}_{c,j} e^{-i\phi})\right)$, with $\theta = \frac{\eta_j}{2}\int_0^t\,dt' g(t')$, which carries out the transformation
\begin{equation}
\hat{D}_j(\theta,\phi) \hat{a}_{c,j} \hat{D}_j^\dagger(\theta,\phi)
= \hat{a}_{c,j} - \theta e^{i\phi} \;.
\end{equation}
While the boson sculpting scheme does not require the use of the displacement operator, it can be useful for tomography when used with the parity gate.

\subsubsection{Arithmetic subtractions of phonons\label{ss:arithmetic-subtraction-implementation}}

The arithmetic subtraction operator on mode $k$ can be performed by adiabatically driving the red sideband transition given in \eqref{eq:rsb-hamiltonian}. This is done by slowly varying the laser detuning and coupling strengths over a time interval $t\in [0,\tau]$ as \cite{Um}
\begin{subequations}
\begin{align}
\delta_1(t) = \delta_2(t) &= -\nu_k + \Delta_0\cos(\tfrac{\pi t}{\tau}) \;,\\ 
g_1(t) = g_2(t) &= g_0\sin(\tfrac{\pi t}{\tau}) \;,
\end{align}
\end{subequations}
where the detuning $\Delta_0 =\tfrac{1}{2}\sqrt{n_\text{max} + 1}\eta_k g_0$ depends on $n_\text{max}$, the maximum number of phonons to be subtracted. When $\tau$ is large, the adiabatic theorem states that the eigenstates of the initial Hamiltonian evolves to the corresponding eigenstates of the final Hamiltonian with the same eigenvalue \cite{adiabatic-theorem}. In this case, this results in an adiabatic transfer of the states from $\ket{g}\ket{n}_{c,k} \to \ket{e}\ket{n-1}_{c,k}$, given that the adiabatic condition $\tau \gg 1/g\eta_k$ is met \cite{Um}. A final carrier transition is performed to reset the spin to $\ket{g}\ket{n-1}_{c,k}$. As the red sideband transition does not affect the state $\ket{g}\ket{0}_{c,k}$, this final transition excites the spin to the state $\ket{e}\ket{0}_{c,k}$. A projective measurement on the spin thus removes the vacuum component of the state, completing the arithmetic subtraction process.

\subsubsection{Beam-Splitting operations between the collective modes\label{sec:BS}}

In recent years, significant progress have been made studying the motional degree of freedom of trapped ions \cite{Wentao}. This has led to the demonstration of new tools and techniques that allow for better control over such systems. For instance, due to Coulomb interactions between ions, a beam-splitter-like coupling between the motional modes of two trapped ions has been experimentally demonstrated \cite{Toyoda}. This has led to many applications, such as the implementation of the controlled-SWAP gate for machine learning algorithms \cite{Nguyen,nguyen2021} and the study of quantum walks using phonons \cite{Masaya}. 

The beam-splitter transformation is a second order operation that can be performed by setting $\delta_1=-\delta_2=\nu_j-\nu_k$ and $\phi_1=-\phi_2=\pi-\phi$, so that
\begin{equation}\label{eq:bs-hamiltonian}
\hat{H}_{B,j,k}(t)
= \hbar\hat{\sigma}_x \frac{g(t)\eta_j\eta_k}{2}\left(
    \hat{a}^{\dagger}_{c,j}\hat{a}_{c,k} e^{i\phi} + \hat{a}_{c,j}\hat{a}^{\dagger}_{c,k} e^{-i\phi}
\right) \;.
\end{equation}
By setting the state of the spin to $\vert-\rangle = \frac{1}{\sqrt{2}}\left(\vert e \rangle-\vert g \rangle\right)$, the time evolution is exactly the unitary
\begin{equation}\label{eq:BS-unitary}
    \hat{B}_{j,k}(\theta, \phi) = \exp\left[i\frac{\theta}{2} \left(\hat{a}_{k}^{\dagger}\hat{a}_{j}e^{i\phi}+\hat{a}_{k}\hat{a}_{j}^{\dagger}e^{-i\phi}\right) \right] \;,
\end{equation}
with $\theta = \eta_j\eta_k\int_0^t\,dt' g(t')$. This carries out the beam-splitting transformations
\begin{equation*}
    \hat{B}_{j,k}(\theta,\phi)\hat{a}_{c,j}\hat{B}_{j,k}^\dagger(\theta,\phi)
        = \cos(\scriptstyle\case{\theta}{2}) \hat{a}_{c,j} -
        i e^{i\phi} \sin(\scriptstyle\case{\theta}{2}) \hat{a}_{c,k} \;,
\end{equation*}
\begin{equation*}
    \hat{B}_{j,k}(\theta,\phi)\hat{a}_{c,k}\hat{B}_{j,k}^\dagger(\theta,\phi)
        = \cos(\scriptstyle\case{\theta}{2}) \hat{a}_{c,k} -
        i e^{-i\phi} \sin(\scriptstyle\case{\theta}{2}) \hat{a}_{c,j} \;.
\end{equation*}
As we will see in the following section, the beam-splitting operation is vital for preparing the initial state in the boson sculpting scheme. Furthermore, the parity gate can be performed by choosing different values for the $\phi_1$ and $\phi_2$, as detailed in \cite{Gan}. Together with the displacement operation covered in the previous section, a direct measurement of the joint Wigner function can be performed \cite{Bishop}.

\section{Study of the effects of heating and decoherence}\label{section:noise}

Now that we have defined the sequence of operations for our scheme, we simulate its experimental implementation while considering realistic noise conditions in this section. Here, we assume two sources of external perturbation: the damping of the motional modes due to fluctuations of the trap frequency, and the heating of the motional modes. We model the first effect by an interaction with a phase damping reservoir, and the second with an amplitude damping reservoir. Then, the time evolution of the quantum state is computed via the master equation \cite{PRA2000}

\begin{widetext}
\begin{eqnarray}\label{mastereq}
    \frac{d\hat{\rho}(t)}{dt}  &=& \frac{-i}{\hbar} \Big[\hat{\mathcal{H}}_{\mathrm{I}}(t),\hat{\rho} (t)\Big] + \sum_{r=1}^4 \Bigg( \frac{\gamma_{r}}{2}(\bar{n}+1) (2 \hat{a}_{c,r}\hat{\rho}\hat{a}^{\dagger}_{c,r} - \{\hat{a}_{c,r}^{\dagger}\hat{a}_{c,r},\hat{\rho}\}) + \nonumber \\
     &+& \frac{\gamma_{r} \bar{n}}{2} (2
    \hat{a}_{c,r}^{\dagger}\hat{\rho}\hat{a}_{c,r} - 
    \{\hat{a}_{c,r}\hat{a}^{\dagger}_{c,r},\hat{\rho}\}) + \frac{\kappa_{r}}{2} (2 \hat{a}^{\dagger}_{c,r}\hat{a}_{c,r}\hat{\rho}\hat{a}_{c,r}\hat{a}^{\dagger}_{c,r} - \{\hat{a}_{c,r}\hat{a}^{\dagger}_{c,r}\hat{a}_{c,r}^{\dagger}\hat{a}_{c,r},\hat{\rho}\}) \Bigg) \;,
\end{eqnarray}
\end{widetext}
where $\{\cdot,\cdot\}$ is the anticommutator, $\gamma_{r}$ and $\kappa_{r}$ are the decay rates due to the coupling to the amplitude- and phase-damping reservoirs of the r$^{th}$ motional mode respectively, and $\bar{n} = 10^{6}$ is the average number of phonons in the reservoir.

For a given on-resonant transition, the angle of rotation is proportional to the pulse area $\int_0^{\tau}\dd{t}g(t)$. Typically, to mitigate undesired dynamics involving other transitions, the applied pulse shapes have smooth rising and falling edges. In our numerical simulations, we chose a soft-edged square pulse of the form
\begin{equation}\label{eq:couplingstrength}
    g(t;g_0,\tau,t_r) = g_0\times\begin{cases}
    \sin^2(\frac{\pi t}{2t_r}) & \text{for}~ 0 \leq t < t_r \;,\\
    1 & \text{for}~ t_r \leq t \leq \tau-t_r \;, \\
    \sin^2(\frac{\pi (\tau-t)}{2t_r}) & \text{for}~ \tau-t_r < t \leq \tau \;,\\
    0 & \text{otherwise} \;.
    \end{cases}
\end{equation}
The area of the pulse is $\int_0^{\tau}\dd{t}g(t;g_0,\tau,t_r) = g_0(\tau-t_r)$, where $g_0$ is the amplitude and $t_r$ the rise time of the pulse. The pulse amplitude depends on the power output of the laser, where we used $g_0 = \pi/0.004$ \cite{Nguyen}. Meanwhile, a longer rise time reduces the noise due to off-resonant transitions at the cost of a longer pulse time. Unless otherwise stated, we use $t_r/\tau = 0.125$ in our numerical simulations.

The exact expression for the Hamiltonian used in all numerical simulations is given by (\ref{eq:full-hamiltonian}). The trap parameters are $(\omega_x,\omega_z) = 2\pi \times (1.270,0.519)\;\mathrm{MHz}$, with the collective mode frequencies $(\nu_1,\nu_2,\nu_3,\nu_4) = 2\pi \times (1.270,1.159,0.982,0.702)\;\mathrm{MHz}$, and the Lamb Dicke parameters are $(\eta_1,\eta_2,\eta_3,\eta_4) =  (0.067,0.067,0.076,0.094)$. The numerical solutions of the above equation were computed using the QuantumOptics.jl framework \cite{QuantumOptics}.  

\begin{figure}[t!]
\begin{center}
    \includegraphics[width=0.48\textwidth]{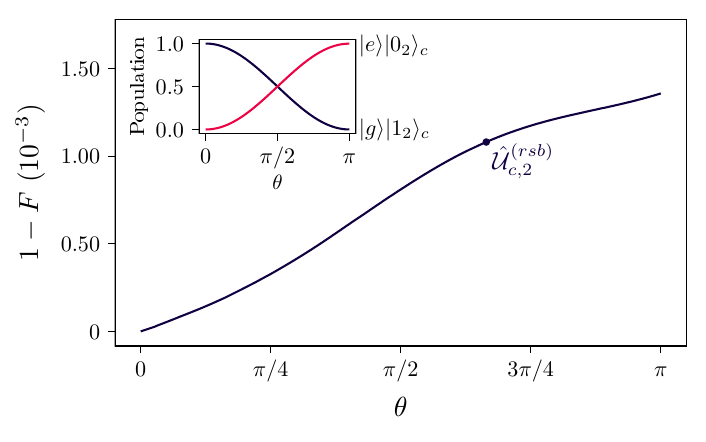}
\end{center}
\vspace{-0.8cm}
    \caption{Red sideband transition. Main plot: infidelity of the target state against the angle of rotation, $\theta$, acting on the $2^\text{nd}$ collective mode. Inset: population of the states $\ket{e}\ket{0_2}_c$ and $\ket{g}\ket{1_2}_c$ for the same range of $\theta$.}\label{simu:RSB}
\end{figure}

\subsection{Numerical simulation of the gates}

\begin{figure}[b!]
\begin{center}
    \includegraphics[width=0.48\textwidth]{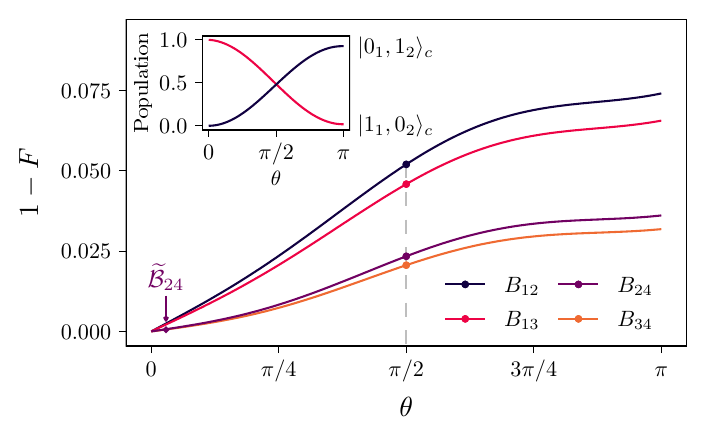}
\end{center}
\vspace{-0.8cm}
    \caption{Beam-splitting transformation between motional modes. The detunings are set to $\delta_1 = \nu_j-\nu_k$ and $\delta_2 = -\pqty{\nu_j-\nu_k}$ for the transformation \eqref{eq:BS-unitary}. Main plot: infidelity against angle of rotation $\theta$, acting on pairs of collective modes. Inset: population of states $\ket{0_1,1_2}_c$ and $\ket{1_1,0_2}_c$ for the same range of $\theta$.
    }\label{simu:BS}
\end{figure}

\begin{figure}[t]
\begin{center}
    \includegraphics[width=0.48\textwidth]{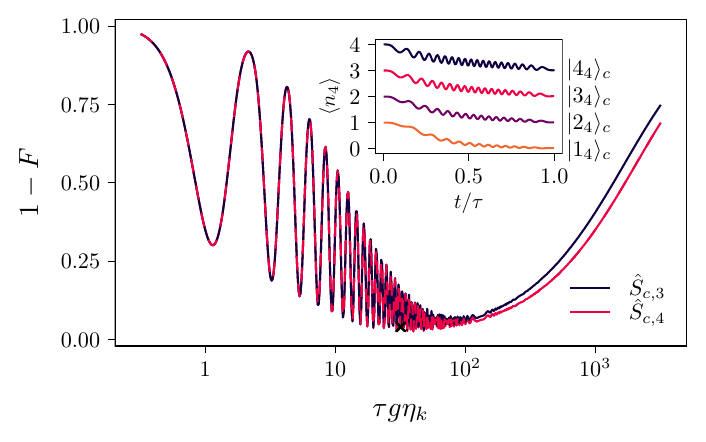}
\end{center}
\vspace{-0.8cm}
    \caption{Adiabatic subtraction from the $3^{rd}$ and $4^{th}$ collective modes. Main plot: infidelity against the dimensionless scaled gate duration $\tau g\eta_k$. An adiabatic passage requires $\tau \gg 1/g\eta_k$. However, if the pulse duration $\tau$ is too large, decoherence effects become significant. Inset: the average number of phonons plotted against the ratio of time over the pulse duration $\tau$. Initial states are $\ket{1_4}_c$, $\ket{2_4}_c$, $\ket{3_4}_c$, and $\ket{4_4}_c$. $\tau$ is chosen from the optimal point in the main plot (see the black X sign).}\label{simu:S}
\end{figure}

Before considering the entire sequence of operations given in Fig.~\ref{fig:cuircuitsab}, which gives rise to the target state, we shall study the effects of noise on each gate separately. In these numerical simulations, contributions of the off-resonant terms are included by considering the full Hamiltonian \eqref{eq:full-hamiltonian}. Heating and decoherence effects are included using Eq.~\eqref{mastereq}, with decay rates $\bar{n}\gamma_{0_{1}} = 15~\mathrm{phonons}/\mathrm{s}$, $ \bar{n}\gamma_{0_{(2,3,4)}} = 0.675~\mathrm{phonons}/\mathrm{s}$, and $\kappa_0 = 0.075~\mathrm{ms}^{-1}$. To implement each gate, the laser detunings are set according to Table \ref{table:detunings-transformations}.

In Fig.~\ref{simu:RSB}, we plot the infidelity of the sideband transition while varying the pulse length. We address the second collective motional modes in order to get the transformation \eqref{eq:rsb-definition2}, i.e. $\ket{g}\ket{1_2}_c \to \ket{e}\ket{0_2}_c $. For a duration up to a $\pi$-pulse, this gate exhibits a robust behaviour with infidelities bellow $10^{-3}$. The blue dot in Fig.~\ref{simu:RSB} corresponds to the transformation \eqref{eq:rsb-trick}. In the inset, we plot the the population transfer from $\ket{g}\ket{1_2}_c$ to $\ket{e}\ket{0_2}_c$.

\begin{figure*}[t]
    \centering
    \includegraphics[width=\textwidth]{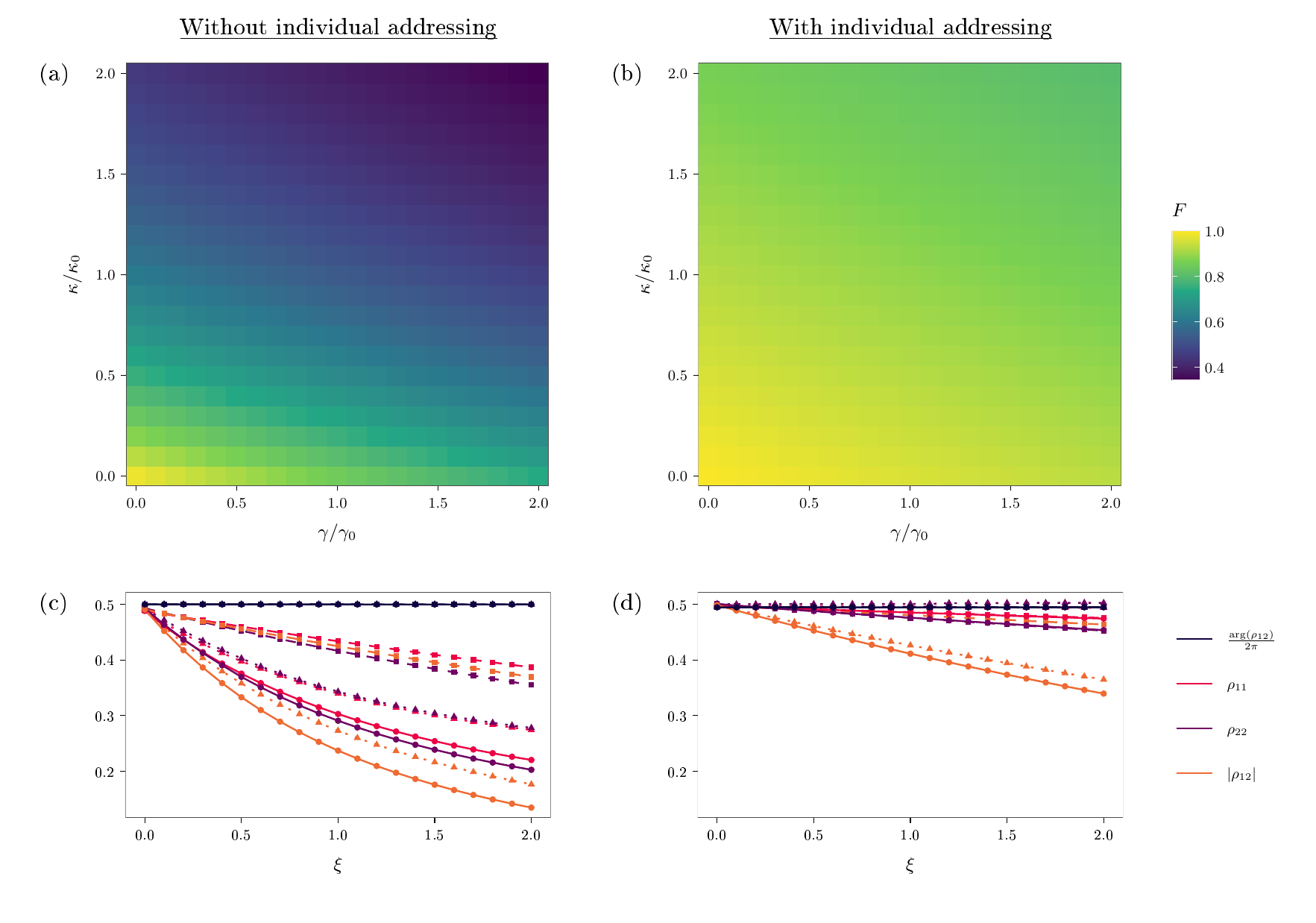}
    \caption{Numerical results for both scenarios in the presence of noise. (a) and (b) Fidelities of the final state against the target state for different coupling strengths to the reservoirs. The reference values of the decay rates are $\bar{n}\gamma_{0_{1}} = 15~\mathrm{phonons}/\mathrm{s}$, $ \bar{n}\gamma_{0_{(2,3,4)}} = 0.675~\mathrm{phonons}/\mathrm{s}$, and $\kappa_0 = 0.075~\mathrm{ms}^{-1}$. (c) and (d) Elements of $\hat{\rho}$, the density matrix of the final state, where $\rho_{11} = \bra{\varnothing} \hat{a}_{c,1}^{}\hat{a}_{c,2}^{} ~\hat{\rho}~ \hat{a}_{c,1}^{\dagger}\hat{a}_{c,2}^{\dagger} \ket{\varnothing}$, $\rho_{22} = \bra{\varnothing} \hat{a}_{c,3}^{}\hat{a}_{c,4}^{} ~\hat{\rho}~ \hat{a}_{c,3}^{\dagger}\hat{a}_{c,4}^{\dagger} \ket{\varnothing}$ and $\rho_{12} = \bra{\varnothing} \hat{a}_{c,1}^{}\hat{a}_{c,2}^{} ~\hat{\rho}~ \hat{a}_{c,3}^{\dagger}\hat{a}_{c,4}^{\dagger} \ket{\varnothing}$. The cases where the system is coupled to both reservoirs ($\kappa/\kappa_0=\gamma/\gamma_0=\xi$) is labelled by the solid lines/circles, to just the heating reservoir ($\gamma/\gamma_0=\xi$, $\kappa=0$) by the dashed lines/squares, and to just the damping reservoir ($\kappa/\kappa_0=\xi$, $\gamma=0$) by the dotted lines/triangles.}
    \label{fig:evolution-with-noise}
\end{figure*}
\begin{table*}
\caption{\label{tab:num_results}Fidelities of the output and implementation times of each gate. The fidelity is computed against the expected state after each step. The duration of each gate is in milliseconds.}
\begin{ruledtabular}
\begin{tabular}{cccccc|ccccccc}
& & \multicolumn{4}{c}{\emph{With Individual Addressing}}  & \multicolumn{7}{c}{Without Individual Addressing}\\
& & $\widetilde{B}_{2,4}$ & $S_3$ & $S_4$ &  RSB$_2$ & $B_{12}$ & $B_{34}$ & $B_{13}$ & $B_{2,4}$ & $S_3$ & $S_4$ & RSB$_2$\\ \hline
    \multirow{2}{*}{Fidelities} & Iso.\footnote{In the isolated case, we correct the state before each gate to estimate the effect of noise on the gates separately.} & 0.988 & 0.863 & 0.960 & 0.997 & 0.938 & 0.860 & 0.742 & 0.849 & 0.863 & 0.960 & 0.997 \\
    & Acum.\footnote{In the accumulated case, the input of each gate is the output of the previous gate.} & --- & 0.860 & 0.912 & 0.891 & --- & 0.809 & 0.611 & 0.530 & 0.525 & 0.622 & 0.534 \\ \hline
    \multirow{2}{*}{Time} & Iso.\footnotemark[1] & 0.02 & 0.54 & 0.44 & 0.02 & 0.52 & 0.56 & 0.79 & 0.37 & 0.54 & 0.44 & 0.02 \\
    & Acum.\footnotemark[2] & --- & 0.56 & 1.00 & 1.02 & --- & 1.08 & 1.87 & 2.24 & 2.78 & 3.22 & 3.24\\ 
\end{tabular}
\end{ruledtabular}
\end{table*}

In Fig.~\ref{simu:BS}, we plot the infidelities of the beam-splitting gate between several pairs of modes. Note that the pairs chosen here are the same pairs discussed in subsection \ref{ss:scenarios}. Addressing these pairs is needed for the implementation of the entangling scheme. Namely, the diamond dot corresponds to the single beam-splitting transformation required for implementing the scenario with individual addressing \ref{ss:scen1}. On the other hand, the circle dots correspond to the 4 beam-splitting transformations required for the implementation of the scenario without individual addressing \ref{ss:scen2}. Since the angle of rotation of any beam-splitting transformation is proportional to the pulse area, larger angles will require longer gate times and hence more errors will accumulate. Consequently, one might predict that the scenario with individual addressing will be more robust against noise. such a prediction will be confirmed in the following subsection. In the inset, we illustrate an example of a beam-splitting transformation between the $1^{\mathrm{st}}$ and $2^{\mathrm{nd}}$ modes with a single phonon initially in the $1^{\mathrm{st}}$ mode. After a $\pi$-pulse, the phonon is transferred to the $2^{\mathrm{nd}}$ with a $\sim 92\%$ fidelity.

In Fig.~\ref{simu:S}, we plot the infidelity of the arithmetic subtraction gate as a function of duration of the adiabatic passage. For this gate, we can see that there is a an optimal time duration for which the infidelity is minimized. If the pulse length is shorter, the fidelity of the gate suffers because the adiabatic passage to too fast. On the other hand, if the duration is longer, the effects of noise become dominant.

\subsection{Numerical simulation of the sculpting scheme}

In Figure~\ref{fig:evolution-with-noise}, we can see that the fidelities in the scenario with individual addressing are considerably more robust against the noise. This can be explained by the fact that the run time for the scenario without individual addressing is almost 3 times longer that the other scenario (cf. Table~\ref{tab:num_results}). For the scenario with individual addressing, the computed fidelity is $0.997$ for the case without noise (bottom left of the color map). For the extreme case with only decoherence, $\kappa = 0$ and $\gamma = 2 \gamma_0$, it is found to be $0.917$. For the case with only heating, $\kappa = 2 \kappa_0$ and $\gamma = 0$, the fidelity is equal to $0.846$. In the upper right corner, where the system is coupled to both baths with $\kappa = 2 \kappa_0$ and $\gamma = 2 \gamma_0$, the fidelity reads $0.778$. In the scenario without individual addressing, following the same order, the fidelities at the corners of the color map are $0.983$, $0.733$, $0.445$, and $0.337$.

In the case where the subtraction operation is performed with the adiabatic blue sideband, we instead obtain the fidelities $0.923$, $0.859$, $0.811$, and $0.754$ at the corners of the color map with individual addressing; and $0.921$, $0.693$, $0.434$, and $0.331$ without individual addressing. When using the blue sideband instead of the red, the behaviour of the system in the presence of noise is similar with slightly lower fidelities.

In subfigures (c) and (d), we plot the entries of the density matrix of the final state as a function of the coupling strength to the reservoirs. Overall, the subtraction scheme is more sensitive to the damping in both scenarios. However, the difference in sensitivity is relatively lower for the scenario with individual addressing. 

In the cases without noise, the fidelities of the final state are not perfect despite the unitary dynamics of our system. This is mainly due to the contributions of the off-resonant terms of the Hamiltonian, as discussed in subsection \ref{ss:Hamiltonian}. By comparing the isolated fidelities to the isolated run times in Table \ref{tab:num_results}, one might expect that the longer the run time of a gate, the more it accumulates such contributions (errors), and the less its isolated fidelity. While this might be intuitive, the cases of $\mathcal{B}_{3,4}$ and $\mathcal{B}_{2,4}$ are an example that such an intuition is not always correct. In this example, $\mathcal{B}_{2,4}$ runs for a shorter time, but the isolated fidelity of $\mathcal{B}_{3,4}$ appears to be higher. This can be explained by the fact that each gate corresponds to a different frequency detuning, ergo the contributions of the remaining terms of the Hamiltonian are always different. These contributions can be suppressed by optimizing the shape of the laser pulses. While we used a soft-edged square pulse for all the operations in this work, identifying the most optimal pulse shape remain an open problem. This will allow for achieving even higher fidelities of the final state.

From Table (\ref{tab:num_results}), one might notice that the isolated fidelities of the subtraction gates are not the same. The first subtraction from mode 3 has an isolated fidelity lower than that of the second subtraction from mode 4. This is because the gate times were optimized for the highest accumulated fidelity after both subtractions, as we are only interested in the resulting state given in equation (\ref{afters3s4}). Therefore, this leads to a lower isolated fidelity of the intermediate state between the two subtractions.

The tomography of such states can be performed in two ways. The first method is to reconstruct the density matrix of the final state by performing projective measurements (see the supplementary material of \cite{Gan}). Using the red sideband transition, the state encoded in the motional degree of freedom can be transferred to the internal degree of freedom, and beam-splitting operations such as $B_{12}(\theta,\phi)$ and $B_{34}(\theta,\phi)$ can be used to perform the necessary rotations. The second method is to reconstruct the joint Wigner function of the final state. As the Wigner function of a state is related to the displaced parity operator, it can be reconstructed by displacing the state and measuring the expectation value of the joint parity gate \cite{Bishop,Wang2016,GaoESwap2019}. For details on the implementation of the displacement and parity gates, see the review \cite{Wentao}.

\section{Conclusions}
This article covered the creation of entanglement between bosonic modes in trapped-ion platforms using arithmetic subtractions, a concept first introduced in \cite{sculpting}.

First, we presented analytically how arithmetic subtractions can transform a separable state in the local basis into a maximally entangled state over the collective modes. Then, we showed that this scheme can be adapted to transform a state prepared in the collective basis into an entangled state in the same basis. Finally, we presented a numerical simulation of an experimental implementation of this scheme assuming two kinds of noise source: the first due to the coupling to a phase-damping reservoir, and the second to an amplitude-damping reservoir. 

In fact, by finding the appropriate beam-splitting operations, this subtraction-based entangling scheme can transform a separable state prepared in any basis (local or collective) into an entangled state in either bases. While we have only considered the case of 4 ions, this scheme can be extended for longer chains. However, since the relations between the local and collective bases for longer chains can only be computed via numerical methods \cite{James}, the adaptation of this scheme will need to be done on a case by case basis. On the other hand, the extension of this protocol to the creation of multipartite-multilevel \textit{GHZ}-like entangled states remains an open problem. 

The type of entangled states that can be generated with this subtraction scheme can be used to entangle the internal degree of freedom of different ions with different electronic structures via the red sideband transition.  Also, these ions can be trapped in separated wells, and therefore moved to different locations. In addition, the state of one ion can be measured by performing the readout on another ion of a different species. This can beneficial for quantum logic spectroscopy and quantum error corrections \cite{home,spectre}.

\section*{Acknowledgements}
This research is supported by the National Research Foundation and the Ministry of Education, Singapore, under the Research Centres of Excellence programme.

\bibliography{refs}

\appendix
\section{\label{apd:ME-PE-Complications}Mode entanglement versus particle entanglement}
In the main text, it was mentioned that the nonzero particle entanglement of the initial state $\ket{\text{sym}_{2n}}$ is converted into mode entanglement through the sequence of subtractions. This distinction between \emph{mode} and \emph{particle} entanglement arises in the discussion of identical bosonic particles \cite{Benatti}, as states appear differently in the first-quantization (particle basis) and second-quantization (occupation number basis) picture. Take, for example, a single-particle basis $\{\ket{\phi_1},\ket{\phi_2},\ket{\phi_3},\ket{\phi_4}\}$. The state $\ket{\text{sym}_{4}}$, which describes a system of four particles with one particle in each single-particle state, is written as
\begin{align*}
    \ket{\text{sym}_{4}} &= \underbrace{\ket{1_{\phi_1},1_{\phi_2},1_{\phi_3},1_{\phi_4}}}_{\text{occupation number basis}} \\
    &= \frac{1}{\sqrt{4!}}\Big(
        \ket{\phi_1} \otimes \ket{\phi_2} \otimes \ket{\phi_3} \otimes \ket{\phi_4}\\ 
    &\qquad\qquad{}+{} \ket{\phi_1} \otimes \ket{\phi_2} \otimes \ket{\phi_4} \otimes \ket{\phi_3} \\ 
    &\qquad\qquad{}+{} \dots \\ 
    &\underbrace{\qquad\qquad{}+{} \ket{\phi_4} \otimes \ket{\phi_3} \otimes \ket{\phi_2} \otimes \ket{\phi_1} \Big)}_{\text{particle basis}},
\end{align*}
where we sum over all possible permutations in the particle basis. Then, the mode entanglement of this system is its entanglement in the occupation number basis, and analogously for its particle entanglement. Here, $\ket{\text{sym}_{4}}$ is clearly separable in the occupation number basis, and hence has zero mode entanglement. However, it is highly entangled in the particle basis.

Although the latter form of entanglement is not directly accessible, the notion of particle entanglement is important as it has been shown that nonzero \emph{particle} entanglement is necessary for extracting \emph{mode} entanglement from a state using only subtraction operations \cite{Morris}. The increase in mode entanglement, accompanied by a reduction in the amount of particle entanglement, leads to the statement that one type of entanglement is converted into the other.

\begin{figure}
    \centering
    \includegraphics{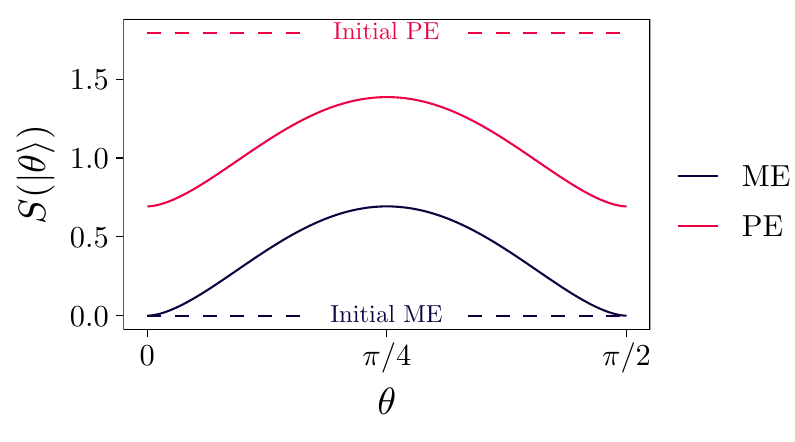}
    \caption{Mode and particle entanglement after the subtraction sequence $\hat{b}_{-\theta}\hat{b}_{+\theta}$. The increase in mode entanglement is accompanied by a decrease in particle entanglement, but the sum $S_{\text{PE}}+S_{\text{ME}}$ is not conserved.}
    \label{fig:A1-PE-ME}
\end{figure}

However, it is an open problem whether this statement can be read as a quantitative interconversion of resources. To illustrate it, we present an example that takes $\ket{\text{sym}_{4}}$ as initial state and involves only pure states. We quantify the particle entanglement $S_{\text{PE}}$ (respectively, mode entanglement $S_{\text{ME}}$) by calculating the von Neumann entropy maximised over all bipartitions of the state in the particle basis (respectively, occupation number basis). Then obviously for the initial state it holds
\begin{equation}
    S_{\text{PE}}(\ket{\text{sym}_{4}}) = \log(6), \qquad S_{\text{ME}}(\ket{\text{sym}_{4}}) = 0.
\end{equation}
We will consider the subtraction sequence $\hat{b}_{-\theta}\hat{b}_{+\theta}\ket{\text{sym}_{4}}$, where $\hat{b}_{+\theta}$ and $\hat{b}_{-\theta}$ are defined as
\begin{equation}
    \hat{b}_{\pm\theta} \equiv \frac{1}{\sqrt{2}}\pqty{
            \sin\theta \; \hat{a}_1
            \pm \cos\theta \; \hat{a}_2\
            + \tfrac{1}{\sqrt{2}}\; \hat{a}_3
            \mp \tfrac{1}{\sqrt{2}}\; \hat{a}_4
        }.
\end{equation}
The resulting state, $\ket{\theta}$, is
\begin{equation}
\begin{aligned}
    \ket{\theta} &\equiv \frac{1}{\abs{\hat{b}_{-\theta}\hat{b}_{+\theta}\ket{\text{sym}_{4}}}}\hat{b}_{-\theta}\hat{b}_{+\theta}\ket{\text{sym}_{4}} \\
    &= \cos\theta \ket{1_{\phi_1}0_{\phi_2}1_{\phi_3}0_{\phi_4}} + \sin\theta \ket{0_{\phi_1}1_{\phi_2}0_{\phi_3}1_{\phi_4}} \\
    &= \cos\theta\frac{1}{\sqrt{2}}\pqty{ 
        \ket{\phi_1}\otimes\ket{\phi_3} + \ket{\phi_3}\otimes\ket{\phi_1}
    } \\
    &\qquad{}+{} \sin\theta\frac{1}{\sqrt{2}}\pqty{ 
        \ket{\phi_2}\otimes\ket{\phi_4} + \ket{\phi_4}\otimes\ket{\phi_2}
    }.
\end{aligned}
\end{equation}
The corresponding values of mode and particle entanglement are plotted in Fig.~\ref{fig:A1-PE-ME}. There is indeed a decrease in particle entanglement accompanying the increase in mode entanglement, but the total amount is clearly not conserved.

The scheme being probabilistic, one might think that the probability of success of the transformation should enter quantitative balances. However, this is unlikely to fix the reported discrepancy: for $\theta\rightarrow 0$, the probability of success tends to $1$; nonetheless, the subtraction has significantly reduced $S_{\text{PE}}$ while $S_{\text{ME}}\approx 0$. Perhaps another of the many measures of ME \cite{Benatti} could capture a quantity that is conserved; or perhaps, no quantitative connection should be sought in a transformation between PE and ME. This remains an open question which might be of interest for further theoretical study. 

\section{\label{apd:general-GHZ-n}Sculpting $\ket{\text{GHZ}_{2n}}$ for general $n$}
In this section, as we will take all operators to be defined in the same basis, the subscripts labelling the collective and local modes will be dropped.

\subsection{GHZ state with ladder operators}
We turn to an alternate subtraction sequence for a $2n$-partite GHZ state adapted from an earlier linear optics protocol \cite{phd-linear-optics}. The subtraction sequence is defined as
\begin{equation}\label{eq:alternate-subtraction-sequence}
\begin{aligned}
    \hat{\mathcal{J}}' 
    &\equiv \frac{1}{2^n}\prod_{j=0}^{n-1}(\hat{a}_{2j\oplus1} - \hat{a}_{2j\oplus2} + \hat{a}_{2j\oplus3} + \hat{a}_{2j\oplus4})\\
    &= \frac{1}{2}(\hat{a}_1-\hat{a}_2+\hat{a}_3+\hat{a}_4)\frac{1}{2}(\hat{a}_3-\hat{a}_4+\hat{a}_5+\hat{a}_6)\cdots\\
    &\qquad{}\times{}\frac{1}{2}(\hat{a}_{2n-3}-\hat{a}_{2n-2}+\hat{a}_{2n-1}+\hat{a}_{2n})\\
    &\qquad{}\times{}\frac{1}{2}(\hat{a}_{2n-1}-\hat{a}_{2n}+\hat{a}_1+\hat{a}_2),
\end{aligned}
\end{equation}
where $x\oplus y := 1 + (x+y-1)\bmod 2n$ is understood to addition cylic over $1,2,\dots,2n$. To work out the action of $\hat{\mathcal{J}}'$ on $\ket{\text{sym}_{2n}}$, consider the annihilation operators $\hat{a}_{2j+1}$ and $\hat{a}_{2j+2}$, which only appear in the two factors,
\begin{widetext}
\begin{align}
    \hat{\mathcal{J}}'\ket{\text{sym}_{2n}}
    &\propto(\cdots)(\hat{a}_{2j-1}-\hat{a}_{2j}+\hat{a}_{2j+1}+\hat{a}_{2j+2})(\hat{a}_{2j+1}-\hat{a}_{2j+2}+\hat{a}_{2j+3}+\hat{a}_{2j+4})(\cdots) \ket{\cdots1_{2j+1}1_{2j+2}\cdots}\nonumber\\\label{eq:ghz-annihilation-state}
    &= (\cdots)\Big(
        (\hat{a}_{2j-1}-\hat{a}_{2j})(\hat{a}_{2j+1}-\hat{a}_{2j+2}) + 
        (\hat{a}_{2j+1}+\hat{a}_{2j+2})(\hat{a}_{2j+3}+\hat{a}_{2j+4})
    \Big)(\cdots)\ket{\text{sym}_{2n}} +\\\nonumber
    &\qquad\qquad(\cdots)\pqty{ (\hat{a}_{2j+1}+\hat{a}_{2j+2})(\hat{a}_{2j+1}-\hat{a}_{2j+2}) }(\cdots) \ket{\text{sym}_{2n}} + \\\nonumber
    &\qquad\qquad(\cdots)\pqty{ (\hat{a}_{2j-1}-\hat{a}_{2j})(\hat{a}_{2j+3}+\hat{a}_{2j+4}) }(\cdots) \ket{\text{sym}_{2n}}.
\end{align}
\end{widetext}
In this expression, the second term $(\hat{a}_{2j+1}+\hat{a}_{2j+2})(\hat{a}_{2j+1}-\hat{a}_{2j+2}) = \hat{a}_{2j+1}^2-\hat{a}_{2j+2}^2$ vanishes as $\ket{\mathrm{sym}_{2n}}$ does not contain any doubly occupied states. Meanwhile, the third term $(\hat{a}_{2j-1}-\hat{a}_{2j})(\hat{a}_{2j+3}+\hat{a}_{2j+4})$ survives the preceding factor ($\hat{a}_{2j-3}-\hat{a}_{2j-2}+\hat{a}_{2j-1}+\hat{a}_{2j}$) and succeeding factor ($\hat{a}_{2j+3}-\hat{a}_{2j+4}+\hat{a}_{2j+5}+\hat{a}_{2j+6}$) only in the form
\begin{align*}
    &(\hat{a}_{2j-3}-\hat{a}_{2j-2})(\hat{a}_{2j-1}-\hat{a}_{2j})\\
    &{}\times{}(\hat{a}_{2j+3}+\hat{a}_{2j+4})(\hat{a}_{2j+5}+\hat{a}_{2j+6}),
\end{align*}
as the other terms introduce double annihilations. Continuing this argument with more preceding and succeeding factors, we are left with the expression
\begin{align*}
    &(\hat{a}_1-\hat{a}_2)(\hat{a}_3-\hat{a}_4)\cdots\\
    &{}\times{}(\hat{a}_{2j-3}-\hat{a}_{2j-2})(\hat{a}_{2j-1}-\hat{a}_{2j}) \\
    &{}\times{}(\hat{a}_{2j+3}+\hat{a}_{2j+4})(\hat{a}_{2j+5}+\hat{a}_{2j+6})\cdots\\
    &{}\times{}(\hat{a}_{2n-3}+\hat{a}_{2n-2})(\hat{a}_{2n-1}+\hat{a}_{2n}).
\end{align*}
This consists of $n-1$ annihilations, and the final annihilation $\hat{a}_{2n-1}-\hat{a}_{2n}+\hat{a}_1+\hat{a}_2$ causes this term to vanish. Hence, only the first term in equation~\eqref{eq:ghz-annihilation-state} survives after performing all $n$ annihilations. Therefore,
\begin{widetext}
\begin{equation}
\begin{aligned}
    \hat{\mathcal{J}}'\ket{\mathrm{sym}_{2n}}
    &\propto \pqty{(\hat{a}_1+\hat{a}_2)(\hat{a}_3+\hat{a}_4)\cdots(\hat{a}_{2n-1}+\hat{a}_{2n})+ (\hat{a}_1-\hat{a}_2)(\hat{a}_3-\hat{a}_4)\cdots(\hat{a}_{2n-1}-\hat{a}_{2n})}\ket{\mathrm{sym}_{2n}}\\
    &\propto \pqty{(\hat{a}_1^\dagger+\hat{a}_2^\dagger)(\hat{a}_3^\dagger+\hat{a}_4^\dagger)\cdots(\hat{a}_{2n-1}^\dagger+\hat{a}_{2n}^\dagger)+ (\hat{a}_2^\dagger-\hat{a}_1^\dagger)(\hat{a}_4^\dagger-\hat{a}_3^\dagger)\cdots(\hat{a}_{2n}^\dagger-\hat{a}_{2n-1}^\dagger)}\ket{\varnothing}.
\end{aligned}
\end{equation}
\end{widetext}
Unlike $\hat{\mathcal{J}}$ defined in the main text, this alternate scheme requires a final sequence of beam-splitters $\overline{\mathcal{B}}$
\begin{equation}\label{eq:alternate-subtraction-sequence-final-beam-splitter}
\overline{\mathcal{B}} \equiv \prod_{j=1}^{n} \mathcal{B}_{2j-1,2j} = \mathcal{B}_{1,2} \mathcal{B}_{3,4} \cdots \mathcal{B}_{2n-1,2n},
\end{equation}
where $\mathcal{B}_{p,q}$ are the same 50-50 beam-splitting operations as in the main text. The sequence $\overline{\mathcal{B}}$ brings the state to
\begin{equation}
\begin{aligned}
    \overline{\mathcal{B}}\hat{\mathcal{J}}'\ket{\mathrm{sym}_n} &\propto \tfrac{1}{\sqrt{2}}\pqty{ \hat{a}_1^\dag \hat{a}_3^\dag\cdots \hat{a}_{2n-1}^\dag + \hat{a}_2^\dag \hat{a}_4^\dag \cdots \hat{a}_{2n}^\dag }\ket{\varnothing} \\
    &\equiv \ket{\mathrm{GHZ}'_{2n}}.
\end{aligned}
\end{equation}
Note that for $n$ even, $\ket{\mathrm{GHZ}'_{2n}}$ differs in a minus sign compared to the state $\ket{\mathrm{GHZ}_{2n}}$ introduced in the main text. In that case, the red sideband transition $\hat{U}_{\text{rsb},j}(\theta=2\pi)$ on any $j$ brings $\ket{\mathrm{GHZ}'_{2n}}$ to $\ket{\mathrm{GHZ}_{2n}}$ and vice versa.

\subsection{GHZ state with arithmetic subtraction}
Firstly, we rewrite the operator $\overline{\mathcal{B}}\hat{\mathcal{J}}'$ as
\begin{equation}
\begin{aligned}
    \overline{\mathcal{B}}\hat{\mathcal{J}}'
    &= \pqty{\prod_{i=1}^{n}\mathcal{B}^\dagger_{2i,2i \oplus 1}} \pqty{\prod_{j=1}^{n}\hat{a}_{2j-1}} \\
    &\qquad{}\times{}\pqty{\prod_{k=1}^{n}\mathcal{B}_{2k,2k\oplus 1}} \pqty{\prod_{l=1}^{n}\mathcal{B}_{2l-1,2l}}.
\end{aligned}
\end{equation}
As before, we replace $\hat{a}_{2j-1}$ with the arithmetic subtraction operator $S_{2j-1} = \pqty{\hat{a}_{2j-1}^\dag \hat{a}_{2j-1}+1}^{-\frac{1}{2}}\hat{a}_{2j-1}$, where we obtain
\begin{align}\nonumber
    &\ket{\psi_{\text{f},2n}}\\\nonumber
    &\propto \pqty{\prod_{i=1}^{n}\mathcal{B}^\dagger_{2i,2i \oplus 1}}
    \pqty{\prod_{j=1}^{n}S_{2j-1}}\\\nonumber
    &\qquad{}\times{}\pqty{\prod_{k=1}^{n}\mathcal{B}_{2k,2k\oplus 1}} \pqty{\prod_{l=1}^{n}\mathcal{B}_{2l-1,2l}}\ket{\text{sym}_{2n}} \\
    &\propto
    \pqty{\prod_{i=1}^{n}\mathcal{B}^\dagger_{2i,2i \oplus 1}}
    \pqty{\prod_{j=1}^{n} \frac{1}{\sqrt{\hat{a}_{2j,2j\oplus 1}^\dagger \hat{a}_{2j,2j\oplus 1} + 1}}}\\\nonumber
    &\qquad{}\times{}\pqty{\prod_{i=1}^{n}\mathcal{B}_{2i,2i \oplus 1}}
    \ket{\text{GHZ}'_{2n}} \\\nonumber
    &\propto \frac{\pqty{\sqrt{2}-1}^n}{\sqrt{2^{n+1}\pqty{3^n+1}}}\Bigg(
    \bigotimes_{j=1}^n\pqty{
        \ket{1_{2j}0_{2j\oplus1}} +
        \pqty{\sqrt{2}+1}^2\ket{0_{2j}1_{2j\oplus1}}
    } \\\nonumber
    &\qquad {}+{} \bigotimes_{j=1}^n\pqty{
        \pqty{\sqrt{2}+1}^2\ket{1_{2j}0_{2j\oplus1}} +
        \ket{0_{2j}1_{2j\oplus1}}
    }\Bigg).
\end{align}
Compared to the target state, this state has the fidelity
\begin{equation*}
\abs{\braket{\psi_{\text{f},2n}}{\text{GHZ}'_{2n}}}^2 = \frac{\pqty{\sqrt{2}-1}^n+\pqty{\sqrt{2}+1}^n}{\sqrt{2^n\pqty{3^n+1}}},
\end{equation*}
and the success probability $(3^n+1)/2^{3n-1}$.

Like before, we can use the red sideband trick to obtain the maximally entangled state. Comparing the state of the system right after the subtraction process,
\begin{widetext}
\begin{align*}
&\pqty{\prod_{j=1}^{n}\hat{a}_{2j-1}}
    \pqty{\prod_{k=1}^{n}\mathcal{B}_{2k,2k\oplus 1}} \pqty{\prod_{l=1}^{n}\mathcal{B}_{2l-1,2l}}\ket{\text{sym}_{2n}}\\
    &\qquad\qquad{}\propto{} \bigotimes_{j=1}^n\pqty{
        \ket{1_{2j}0_{2j\oplus1}} +
        \ket{0_{2j}1_{2j\oplus1}}
    } + \bigotimes_{j=1}^n\pqty{
        \ket{1_{2j}0_{2j\oplus1}} -
        \ket{0_{2j}1_{2j\oplus1}}
    },\\
&\pqty{\prod_{j=1}^{n}S_{2j-1}}
    \pqty{\prod_{k=1}^{n}\mathcal{B}_{2k,2k\oplus 1}} \pqty{\prod_{l=1}^{n}\mathcal{B}_{2l-1,2l}}\ket{\text{sym}_{2n}} \\
    &\qquad\qquad{}\propto{} \bigotimes_{j=1}^n\pqty{
        \sqrt{2}\ket{1_{2j}0_{2j\oplus1}} +
        \ket{0_{2j}1_{2j\oplus1}}
    } + \bigotimes_{j=1}^n\pqty{
        \sqrt{2}\ket{1_{2j}0_{2j\oplus1}} -
        \ket{0_{2j}1_{2j\oplus1}}
    },
\end{align*}
\end{widetext}
we note that they differ in an extra factor of $\sqrt{2}$ that appears only in the terms where the even modes are occupied. Hence, the exact gate sequence for generating the $\ket{\text{GHZ}_{2n}}$ state with the red sideband correction is
\begin{widetext}
\begin{equation}\label{eq:general-subtraction-scheme}
\begin{aligned}
    \ket{\mathrm{GHZ}'_{2n}} &\propto \pqty{\prod_{h=1}^{n}B_{2h,2h \oplus 1}} \pqty{\prod_{i=1}^{n}\bra{g}U_{\text{rsb},2i}(\tfrac{\pi}{4})}
    \pqty{\prod_{j=1}^{n}S_{2j}}
    \pqty{\prod_{k=1}^{n}B_{2k,2k\oplus 1}^\dagger} \pqty{\prod_{l=1}^{n} B_{2l-1,2l}^\dagger}\ket{\mathrm{sym}_{2n}}.
\end{aligned}
\end{equation}
\end{widetext}
Here, $\bra{g}U_{\text{rsb},2i}\!\pqty{\tfrac{\pi}{4}}$ is to be understood as a red sideband transition on the $(2i)$-th motional mode with $\theta = \frac{\pi}{4}$, followed by a post-selection on the ground state of the internal degree of freedom of the ion. This introduces a factor of $1/\sqrt{2}$ to the even modes, thus correcting the extra factor of $\sqrt{2}$. The success probability with the red sideband correction is $2^{-(2n-1)}$.

Considering that $2n$ red sideband transitions are required to prepare the $\ket{\text{sym}_{2n}}$ from the motional ground state, this means that a total of $3n$ first-order gates (red sideband) and $4n$ second-order gates (beam-splitting and subtraction operations) are required to prepare a $\ket{\text{GHZ}'_{2n}}$ state for general $n$. 

When considering the total operation time of the general scheme, note that if we coupled lasers to $n$ of the $2n$ ions in the ion chain, each beam splitter sequence ($\prod_{l=1}^{n} B_{2l-1,2l}^\dagger$, $\prod_{k=1}^{n}B_{2k,2k\oplus 1}^\dagger$, and $\prod_{h=1}^{n}B_{2h,2h \oplus 1}$) can be performed in parallel. In that case, the time taken for \emph{all} the beam splitter operations in equation~\eqref{eq:general-subtraction-scheme} is $3\times\text{(time taken for one beam splitter operation)}$.
\end{document}